%% file: main.tex
\documentclass[conference]{IEEEtran}
\IEEEoverridecommandlockouts
\usepackage{cite}
\usepackage{amsmath,amssymb,amsfonts}
\usepackage{algorithmic}
\usepackage{graphicx}
\usepackage{textcomp}
\usepackage{xcolor}
\usepackage{siunitx}
\usepackage{comment}
\usepackage{tikz}
\usepackage{amsmath}
\usetikzlibrary{calc}
\usepackage[colorinlistoftodos]{todonotes}

\input{macros}
\def\BibTeX{{\rm B\kern-.05em{\sc i\kern-.025em b}\kern-.08em
    T\kern-.1667em\lower.7ex\hbox{E}\kern-.125emX}}
\begin{document}

\title{Partially Observable Residual Reinforcement Learning for PV-Inverter-Based Voltage Control in Distribution Grids\\

}

\author{\IEEEauthorblockN{Sarra Bouchkati\IEEEauthorrefmark{1},
Ramil Sabirov\IEEEauthorrefmark{1},
Steffen Kortmann\IEEEauthorrefmark{1},
Andreas Ulbig\IEEEauthorrefmark{1}\IEEEauthorrefmark{2} }
\IEEEauthorblockA{\IEEEauthorrefmark{1} IAEW at RWTH Aachen University, Aachen, Germany\\ s.bouchkati@iaew.rwth-aachen.de}
\IEEEauthorblockA{\IEEEauthorrefmark{2} Center Digital Energy, Fraunhofer FIT, Aachen, Germany}
}

\maketitle

\input{00_Abstract}
\input{01_Introduction}

\input{02_Relted_Work}

\input{03_Background}
\input{04_Methodology}
\input{05_Experiments}
\input{06_Conclusion}

\bibliographystyle{IEEEtran}
\bibliography{Bibliography}

\end{document}

%% file: macros.tex
\usepackage{amsfonts} 
\usepackage{hyperref}

\renewcommand{\S}{\mathcal{S}}
\newcommand{\A}{\mathcal{A}}
\newcommand{\s}{\mathbf{s}}
\newcommand{\Obs}{\Omega}
\newcommand{\R}{R}

\renewcommand{\a}{\mathbf{a}}
\renewcommand{\o}{\mathbf{o}}
\newcommand{\E}{\mathbb{E}}
\newcommand{\M}{\mathcal{M}}

\newcommand{\rl}{\mathrm{RL}}
\newcommand{\ctrl}{\mathrm{CT}}
\newcommand{\mix}{\mathrm{mix}}
\newcommand{\maximum}{\mathrm{max}}
\newcommand{\inj}{\mathrm{inj}}

\newcommand{\arl}{\a^\rl}
\newcommand{\actrl}{\a^\ctrl}
\newcommand{\amix}{\a^\mix}
\newcommand{\pirl}{\pi^\rl}
\newcommand{\pictrl}{\pi^\ctrl}

%% file: 00_Abstract.tex
\begin{abstract}

This paper introduces an efficient Residual Reinforcement Learning (RRL) framework for voltage control in active distribution grids. Voltage control remains a critical challenge in distribution grids, where conventional Reinforcement Learning (RL) methods often suffer from slow training convergence and inefficient exploration. To overcome these challenges, the proposed RRL approach learns a residual policy on top of a modified Sequential Droop Control (SDC) mechanism, ensuring faster convergence. Additionally, the framework introduces a Local Shared Linear (LSL) architecture for the Q-network and a Transformer-Encoder actor network, which collectively enhance overall performance. Unlike several existing approaches, the proposed method relies solely on inverters' measurements without requiring full state information of the power grid, rendering it more practical for real-world deployment. Simulation results validate the effectiveness of the RRL framework in achieving rapid convergence, minimizing active power curtailment, and ensuring reliable voltage regulation.

\end{abstract}


%% file: 01_Introduction.tex
\section{Introduction}
With the increasing integration of Distributed Energy Resources (DERs), such as photovoltaic (PV) systems, distribution grids are experiencing higher utilization of grid assets, driven by the intermittent and decentralized nature of renewable energy generation. This transition poses new challenges for grid operators, particularly in maintaining reliable and efficient grid operation.

One of the most critical challenges introduced by DERs is voltage regulation. The volatility and uncertainty associated with Renewable Energy Sources (RES) lead to fluctuations in voltage levels, which can compromise operational reliability \cite{RazaviIntro}, \cite{SHARMA2020Intro}. Effective voltage control is therefore essential to ensure stable and efficient grid operation while maximizing the penetration of RES.

Several strategies have been proposed to address these challenges. Local approaches, such as Q-V droop control \cite{AlamQV15} and P-V droop control \cite{HAQUE2017PV}, rely solely on local measurements and operate without the need for a communication infrastructure. These methods are simple and scalable but generally result in suboptimal solutions, as they lack coordination between distributed resources.
Model-based optimization approaches, on the other hand, offer global optimal solutions and have been extensively used for voltage regulation \cite{Zhang2020Central}, \cite{EVANGELOPOULOS201695}. However, these methods often require an accurate model of the distribution grid and extensive measurement data from all grid participants. However, fulfilling these requirements in practice is challenging due to uncertainties in the distribution grid topology, and the limited deployment of intelligent metering systems.
In contrast to these methods, model-free approaches, such as Deep Reinforcement Learning (DRL), have emerged as promising alternatives for voltage control \cite{Chen_2022}. RL techniques can learn optimal policies directly from data, bypassing the need for an explicit grid model or full state information. This feature makes them particularly well-suited for the volatile and uncertain environment of active distribution grids.

%% file: 02_Relted_Work.tex
\subsection{Related Work}
To this end, several DRL methods for voltage control have been proposed \cite{WangCentral19}, \cite{AlSaffarCentral20}. However, these approaches typically rely on centralized execution, which imposes a high communication burden. In large-scale distribution grids, this heavy reliance on communication infrastructure presents significant challenges, such as data transmission delays, communication bottlenecks, and potential outages. Furthermore, the limited measurement deployment can lead to incomplete or outdated data, compromising the effectiveness of centralized control strategies.

To handle the communication burden inherent in centralized approaches, Multi-Agent Reinforcement Learning (MARL) has been explored, where multiple agents are assigned control responsibilities for specific inverters or sections of the grid. For instance, the framework proposed in \cite{Cao21} employs centralized training with decentralized execution, enabling agents to learn collaboratively, while operating independently during execution. Similarly, the work in \cite{wangMARL21} formulates voltage constraints as barrier functions within a state-constrained MARL framework to ensure feasible and reliable operation. As the number of agents increases, however, the complexity of centralized training scales exponentially due to the enlarged joint action and state spaces. This leads to slower convergence and increased computational demands, particularly in large-scale power systems. Furthermore, a notable drawback, addressed in \cite{wangMARL21}, is the potential for inexplicable behaviors, which can limit their suitability for safety-critical systems.

To address the issue of safety, several works have focused on modifying the reward function or training framework. For instance, \cite{zhang2020deepreinforcementlearningbased} introduces a switch reward mechanism that prioritizes resolving voltage violations before minimizing power losses. This approach prevents unstable training scenarios where excessive penalty values for violations could disrupt the learning process. Similarly, the method proposed in \cite{ZhangProjection24} projects the agent's actions onto the feasible region, ensuring that only safe actions are executed. While this guarantees safety, it can reduce exploration within the action space, potentially limiting the optimality of the learned policy.

Other approaches leverage the power flow model to guide policy learning. The works in \cite{LiuRobustRL21} and \cite{GAO2022ModelAugmented} train a model to approximate the power flow model, followed by training a DRL agent on the learned model for voltage control. Additionally, \cite{GAO2022ModelAugmented} incorporates a safety layer to ensure constraint satisfaction, while the work in \cite{Hossain23GCN} combines DRL with Graph Convolutional Networks (GCNs) to address robustness to topological changes. The GCN extracts relevant features from the grid's topology, which are then fed to the DRL algorithm to improve its decision-making process.

Although these methods show promise in addressing safety and robustness concerns, three key challenges remain: the solutions may not always achieve optimality, a large amount of data is often required for effective training, and training can be time-consuming, with weak convergence and instability in some cases.

Recently, hybrid RL methods have gained attention by combining control priors with deep RL policies. These hybrid algorithms combine the generalization capabilities and informed behavior of prior controllers with the adaptability of DRL for solving complex nonlinear problems. The concept of residual RL, as introduced in \cite{johannink2018residualreinforcementlearningrobot} and\cite{silver2019residualpolicylearning}, builds on this idea by learning a residual policy on top of a fixed control prior. This approach enhances data efficiency and robustness to uncertainties.

The first application of residual RL to the voltage control problem was proposed in \cite{liu2024residualdeepreinforcementlearning}. This method employs model-based optimization as a control prior and trains a residual RL agent to learn a policy within a reduced action space. However, the method requires full information about active and reactive power injections and voltage measurements at all grid buses, which is often impractical in real-world scenarios. \\
The work in \cite{liu2024robustdeepreinforcementlearning} identifies the problem of the limited measurement deployment in active distribution grids, and  introduces a conservative critic to estimate the state-action value function based on the limited measurements. However, the accuracy of this estimation is crucial for the performance of the DRL algorithm, and the learned policy may not be optimal.

\subsection{Contributions}
To address the limitations mentioned above, we propose a novel residual partially-observable RL (RRL) framework for the coordinated control of PV inverters. The key contributions of this work are as follows:

\begin{itemize}
    \item 
    The proposed RRL framework enhances voltage control by learning a residual policy on top of the Sequential Droop Control (SDC) mechanism, facilitating faster convergence and improved training and data efficiency.

    \item 
   We propose a Local Shared Linear (LSL) architecture for the Q-network, inspired by the pattern-capturing capabilities of Convolutional Neural Networks (CNNs), along with a Transformer Encoder-based model for the actor network. This combination effectively captures generalizable patterns, resulting in enhanced learning stability and improved performance.
    
    \item We design a partially-observable framework, which operates solely using the inverters' measurements, eliminating the need for complete state information. 
    
\end{itemize}

%% file: 03_Background.tex
\section{Background}
This section introduces the background of the voltage control problem and residual reinforcement learning. 

\subsection{Voltage Control}
The increasing integration of PV systems in low-voltage (LV) distribution grids is anticipated to cause frequent overvoltage issues. In LV grids, which typically exhibit a high R/X ratio, active power plays a crucial role in voltage regulation. Nevertheless, reactive power remains essential for effective voltage control and minimizing active power curtailment.

\input{Figures/SDC}

In this work, we adopt a modified version of the SDC mechanism~\cite{MAI2021106931} which combines both Q-V and P-V droop control strategies, as illustrated in Figure \ref{fig:sdc}. In the first phase (1), the Q-V droop control follows the conventional droop approach with a deadband, spanning the interval between $V_1$ and $V_2$. When the voltage surpasses the threshold $V_3$, the reactive power is set to its maximum limit $-\overline{Q}_i$ at inverter $i$. If the voltage continues to rise and surpasses the value $V_4$, depicting the upper voltage value, the P-V droop control is activated. In this phase (2), the active power $P_i$ is curtailed linearly from its maximum value $P_i^{\mathrm{max}}$ to counter further voltage rise. This sequential approach enhances voltage regulation by prioritizing reactive power utilization before resorting to active power curtailment.

\subsection{Reinforcement Learning}
RL is a method for solving sequential decision problems based on the
interaction between an agent and an environment\cite{sutton_reinforcement_2018}. The environment is modeled as a Markov decision process (MDP) that is formally defined by the 6-tuple $(\S, \A, T, R, \gamma, \rho_0)$, with state space $\S$ and action space $\A$. At timestep $t$, the environment transitions from a state $\s_t\in\S$ into a state $\s_{t+1}\in\S$ by applying the action $\a_t\in\A$ according to the probability distribution defined by the transition function $T(\s_{t+1} \mid \s_{t}, \a_{t})$, whereby the initial state $\s_0$ is drawn from the start state distribution $\s_0 \sim \rho_0$. During transition, a scalar reward $r_{t+1}\in \mathbb{R}$ is emitted by the reward function $\R(\s_t, \a_t) = r_{t+1}$. The goal of RL is to learn a policy $\pi(\a_t \mid \s_t)$ that maximizes the expected cumulative reward discounted by $\gamma\in(0,1)$. This results in the objective function
\begin{equation}
    \max_\pi \E_{\pi, \M}\left[ \sum_{t=0}^{\infty} \gamma^t r_{t+1} \right].
\end{equation}
The discounted sum of rewards is referred to as the return and is accumulated along trajectories which are the result of the repeated interaction of the agent with the environment executed by sampling from policy and transition function in alternation.

Oftentimes the agent does not have full access to the environment state, which is formally modeled as a partially observable Markov decision process (POMDP). Here, the agent receives observations $\o_t$ from the observation space $\Obs$, drawn from the probability distribution $O(\o_t \mid \s_t)$ based on the true state $\s_t$.

\subsection{Residual Reinforcement Learning}
In Residual RL \cite{johannink2018residualreinforcementlearningrobot, silver2019residualpolicylearning} the RL agent learns a policy $\pirl(\arl\mid\s_t)$ on top of a prior policy $\pictrl(\actrl\mid\s_t)$. The combined action
\begin{equation}
    \amix_t = \actrl_t + \lambda \arl_t
    \label{eq:mixing_actions}
\end{equation}
is passed to the environment, where $\arl_t \sim \pirl(\cdot\mid\s_t)$, $\actrl_t \sim \pictrl(\cdot\mid\s_t)$, and 
$\lambda$ controls the RL agent's influence relative to the prior policy.
This formulation casts the original RL problem as finding a policy in the residual MDP $\M' = (\S, \A, T', R, \gamma, \rho_0)$ that subsumes the prior policy into the new transition function
\begin{equation}
    T'(\s_{t+1} \mid \s_t, \arl_t) = T(\s_{t+1} \mid \s_t, \actrl_t + \lambda \arl_t).
\end{equation}
Embedding the prior policy into the environment's dynamics can simplify the RL agent's exploration and speed up the training \cite{johannink2018residualreinforcementlearningrobot, silver2019residualpolicylearning}. The prior policy guides the agent with a reasonable starting point, allowing it to focus exploration on improving areas where the prior is suboptimal.

In our case, the prior policy corresponds to the deterministic modified SDC controller.

%% file: Figures/SDC.tex
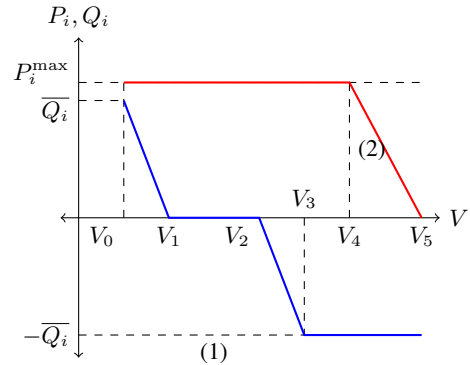
\begin{figure}[ht!]
    \centering
    \begin{tikzpicture}[scale=1.2, every node/.style={font=\small}] 

        \draw[<->] (-0.2,0) -- (4,0) node[right] {$V$};
        \draw[<->] (0,-1.55) -- (0,2) node[above] {$P_i, Q_i$};

        \node[below left] at (0.5,0) {$V_0$};
        \node[below ] at (1,0) {$V_1$};
        \node[below left] at (2,0) {$V_2$};
        \node[above] at (2.5,0) {$V_3$};
        \node[below] at (3,0) {$V_4$};
        \node[below] at (3.8,0) {$V_5$};

        \draw[dashed] (0.5,0) -- (0.5,1.5);
        \draw[dashed] (0,1.3) -- (0.5,1.3);
        \draw[dashed] (2.5,0) -- (2.5,-1.3);
        \draw[dashed] (3,0) -- (3,1.5);
        \draw[dashed] (0,1.5) -- (3.8,1.5) ;
        
        \draw[dashed] (0,-1.3) -- (3,-1.3) ;

        \draw[blue, thick] (0.5,1.3) -- (1,0) -- (2,0) -- (2.5,-1.3) -- (3.8,-1.3);

        \draw[red, thick] (0.5,1.5) -- (3,1.5)-- (3.8,0);

        \node[font=\small] at (1.5,-1.5) {(1)};
        \node[font=\small] at (3.25,0.75) {(2)};

        \node[left] at (0,1.6) {$P_i^{\mathrm{max}}$};
        \node[left] at (0,1.2) {$\overline{Q_i}$};
        \node[left] at (0,-1.3) {$-\overline{Q_i}$};

    \end{tikzpicture}
    \caption{SDC Mechanism for Q-V and P-V Control}
    \label{fig:sdc}
\end{figure}

%% file: 04_Methodology.tex
\section{Approach}
This section presents a detailed overview of the implemented environment for the RRL framework, along with the proposed novel architecture, which integrates an LSL architecture for the Q-network and a transformer encoder-based design for the actor-network.

\subsection{Voltage Control Environment}
We model the distribution grid as a Graph $\mathcal{G} = (\mathcal{N}, \mathcal{E})$, with a set of nodes $\mathcal{N} = \{1, \ldots, n\}$, connected by the edges $\mathcal{E}$. A subset of these nodes have a PV-inverter attached to and represents the set of controllable nodes $M = \{j_1, \ldots, j_m\} \subseteq \mathcal{N}$.   
At timestep $t$, the agent observes the state of the electrical grid through the local information at the nodes the PVs are connected to. In particular, for every controllable node $j\in \M$ the agent receives the local observation
\begin{equation}
    \o_{i,t} = (v_{j,t}, p_{j,t}, q_{j,t}, p_{j}^\maximum),
\end{equation}
where $v_{j,t}$ is the voltage measurement, $p_{j,t}$ and $q_{j, t}$ are the active and reactive power injections, and $p_{j}^\maximum$ describes the maximal possible active power injection for the PV-inverter at timestep $t$. The total observation the agent receives is the concatenation of all local observations, i.e. 
\begin{equation}
    \o_t = (\o_{1,t}, \ldots, \o_{m,t}).
\end{equation}
At each timestep, the agent outputs a 2-dimensional action for each controlled PV-inverter: 
\begin{equation}
    \a_{j,t} = (p^\inj_{j,t}, q^\inj_{j,t}),
\end{equation}
which specifies a desired active and reactive setpoint constrained by the set of feasible points $\mathcal{F}$ of the PV-inverter. As the latter is allowed to adjust both active and reactive power, the set of feasible operating points, as defined in \cite{FORpv}, is given by:
\begin{align}
    \mathcal{F}_{j} := \Bigl\{ 
    &\left( p^\inj_{j}, q^\inj_{j} \right) :
    0 \leq p^\inj_{j} \leq p_{j}^{\max}, \notag \\
    & |q^\inj_{j}| \leq \min\left(\sqrt{s_{j}^{2} - \left(p_{j}^{\inj}\right)^2}, p^\inj_{j}\tan(\phi) \right) 
    \Bigr\},
\end{align}
where $s_j$ denotes the total apparent power of inverter $j$ and the power factor is given by $\cos(\phi)$.\\
To avoid an oscillatory behavior of the different controllers in the environment, the desired setpoint is not applied as-is to the grid but instead applied with a delay $\alpha$, i.e.:
\begin{subequations}
\begin{align}
    p_{j,t+1} &= \alpha \cdot p_{j,t} + (1-\alpha)\cdot p^\inj_{j,t} \label{eq:subeq1}\\
    q_{j,t+1} &= \alpha \cdot q_{j,t} + (1-\alpha)\cdot q^\inj_{j,t}. \label{eq:subeq2}
\end{align}
\end{subequations}
Analogous to the total observation, the total action $\a_t$ fed back to the environment is the concatenation of the individual local actions. The next grid state is obtained by performing a power flow computation, for which we rely on the Pandapower library\footnote{The code for the environment is publicly available at \href{https://github.com/RWTH-IAEW/voltage-control-env}{https://github.com/RWTH-IAEW/voltage-control-env}.}~\cite{pandapower.2018}. 

Similar to previous work~\cite{liu2024residualdeepreinforcementlearning, zhang2020deepreinforcementlearningbased} the reward function consists of two parts. The first penalizes the agent for voltage violations present in the electrical grid
\begin{align}
    R_v(\s_t, \a_t) = - \max_{i\in N} \Bigl[ 
        &\max \left( v_{i,t+1} -  \overline{V}, 0 \right) \notag \\
        &+ \max \left( \underline{V} - v_{i,t+1}, 0 \right)
    \Bigr].
\end{align}

The second incentivizes a maximal use of the available active power generated by the PVs
\begin{equation}
    R_p(\s_t, \a_t) = \begin{cases}
        \frac{1}{|M|}\sum_{j\in M} p_{j, t+1} &\quad \text{if } R_v(\s_t, \a_t) = 0\\
        0 &\quad \text{else}.
    \end{cases}
\end{equation}
The total reward then becomes the sum of both components
\begin{equation}
    R(\s_t, \a_t) = R_v(\s_t, \a_t) +  R_p(\s_t, \a_t).
    \label{eq:reward}
\end{equation}
By restricting rewards to cases where voltage violations are resolved, we ensure that the agent maximizes power utilization without compromising grid safety. A policy that avoids voltage violations strictly dominates one that permits them, preventing the agent from learning unsafe behavior\cite{zhang2020deepreinforcementlearningbased}.

\subsection{Deep Residual Reinforcement Learning}
As the downstream RL learning algorithm in our RRL approach, we use Soft Actor-Critic (SAC)\cite{haarnoja_soft_2018}. It is a state-of-the-art DRL algorithm that embeds the Actor-Critic training architecture into the Maximum Entropy RL framework through entropy regularization.

To deal with the partial observability of the environment, the history of observations is often taken into account by either maintaining a fixed-size sliding window of past observations\cite{mnih_human-level_2015} or using a recurrent neural network architecture\cite{hausknecht2015deep}. Using the RRL approach we find that we can improve upon the control prior without employing either of the two methods. Instead, we train a policy that maps directly from the observations to the actions $\pirl(\a_t\mid\o_t)$, i.e. the fixed-size window of past observations has size \num{1}. For that approach to work reliably, we identify two crucial design choices.
\subsubsection{Observing Controller Actions}
\label{sec:observing-controller-actions}
Although the actions of the control prior are implicitly included in the transition dynamics and do not need to be explicitly observed to learn a policy in the induced residual MDP $\M'$, a selection of prior work still decided to append the actions to the observations of the RL agent\cite{rana_residual_2020, ceola_resprect_2024}. In the Voltage Control Environment, the action space is relatively large compared to the observation space, and appending the actions results in a 50\% increase in the observation space dimension. While this increase in dimensionality can complicate learning and lead to training failures, we find that when paired with an appropriate network architecture, this augmentation results in a better approximation of the Q-function and significantly improves asymptotic performance.

\subsubsection{Network Architecture}
\label{sec:network-architecture}
In many DRL applications, a simple Multi-Layer Perceptron (MLP) is employed for the approximation of the Q-function\cite{WangCentral19, liu2024residualdeepreinforcementlearning, ceola_resprect_2024}. In our work, we found that leveraging the structured nature of the observations and actions in the Voltage Control Environment, which consists of concatenated information of the same kind, leads to significant improvements in training stability and performance. To exploit this structure, we employ a hierarchical feature extraction by prepending shared local linear layers to the MLP (see Figure \ref{fig:q-net-architecture}).
\input{Figures/Network-Architecture}
The perceptive field of one such linear layer pertains to the local information at a single bus, i.e. observations and actions of RL-agent and controller. By stacking these layers, we introduce a mechanism that detects local patterns and computes meaningful local features. Sharing the feature extractor across the entire input enables efficient reusability of these features and forces the network to learn general patterns relevant to all inputs, helping to mitigate overfitting.
This approach is akin to how CNNs operate in tasks such as image classification, where shared filters detect local features across spatial regions\cite{krizhevsky2012imagenet}. Similarly, each neuron in the shared linear layers (shown in color in Figure \ref{fig:q-net-architecture} ) acts like a CNN filter, applied across the input to capture generalizable patterns.

The training performance proved comparably insensitive to the choice of Actor-network architecture. However, we achieved the best results by also exploiting the repeated structure of the inputs, utilizing a Transformer Encoder model\cite{vaswani2017attention}. In our approach, the local information at each bus (i.e., observations and controller actions) is treated as a single input token. These tokens are linearly projected to the appropriate dimension through a shared embedding layer, with a learned positional encoding added\cite{vaswani2017attention}. The tokens are then passed through the encoder consisting of a sequence of attention blocks. For each input token, the encoder generates an output token, which is interpreted as the action of a single PV.
The multi-head attention mechanism within the attention blocks enables efficient message passing by allowing the PVs to attend to each other’s inputs, thus supporting a well-coordinated control effort.

%% file: Figures/Network-Architecture.tex
\begin{figure}[ht!]
    \centering
    \includegraphics[scale=1]{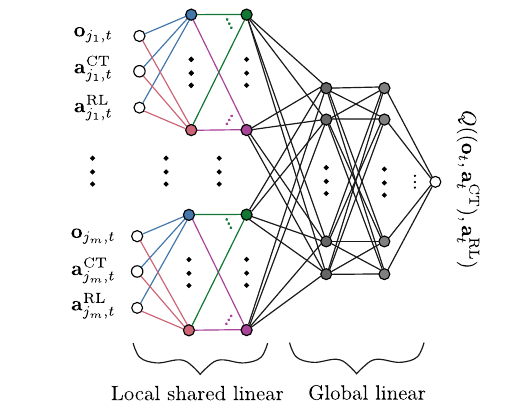}
    \caption{Q-network architecture with one hidden local shared linear layer and two hidden global linear layers.}
    \label{fig:q-net-architecture}
\end{figure}

%% file: 05_Experiments.tex
\section{Experiments}
\subsection{Experimental Setup}
We conduct our experiments on a low-voltage rural distribution grid, with high PV penetration, from Simbench (see Figure \ref{fig:electrical-grid}) \cite{meinecke2020simbench}. \input{Figures/Grid} The grid consists of 128 buses and 27 connected PV systems. Simbench also provides time series data at a 15-minute resolution over one year, with each data point representing a network scenario, i.e., load power demands and PV system nominal power. We exclude data points with minimal active power output (e.g., during the night) and augment scenarios with upper voltage bound violations by perturbing the data. After removing infeasible samples, the final training dataset contains 9,430 data points, of which $\approx\num{90}\%$ represent upper voltage band violations at full power injection.
The voltage bounds are set to $0.95$ p.u. and $1.05$ p.u.

All algorithms use the same neural network architecture as described in \ref{sec:network-architecture}. For the residual approach we also include the control actions in the RL agent observation space (ref. \ref{sec:observing-controller-actions}). We run the training for \num{900}k timesteps or \num{45}k episodes. More detailed hyperparameters regarding the network architecture and training can be seen in Table \ref{tab:hyperparameters}. At the beginning of each new episode, we apply one random scenario from the data set to the distributional grid. To diversify training, the initial setpoint of the PV systems is drawn at random from its feasible region. Every episode runs for 20 time steps before it gets truncated\cite{pardo2018Time}.

\input{Figures/Hyperparameter}

To evaluate the performance of the different algorithms, we use a fixed set of 30 scenarios throughout training. Every 250 episodes, we perform a greedy evaluation run and record the mean cumulative reward over the episodes  (referred to as cumulative reward in the following sections).
Post-training we evaluate the final model of the algorithms on a test dataset containing 1000 unseen scenarios, that we generate by perturbing randomly sampled data points from the training dataset. Here, the initial setpoint of the episodes is set to maximal power injection. We repeat each experiment on 5 different random seeds, and report each metric as the mean over the seeds with its \num{95}\% confidence interval.
Our implementation of the algorithms and the experiments is made publicly available\footnote{Code available at \href{https://github.com/RWTH-IAEW/RRL-voltage-control}{https://github.com/RWTH-IAEW/RRL-voltage-control}}.

For the baseline SDC controller,  we adopt the threshold values $V_0= 0.97 \text{ p.u.}, V_1= 0.99 \text{ p.u.}, V_2= 1.01 \text{ p.u.}, V_3= 1.03 \text{ p.u.},  V_4= 1.045 \text{ p.u. , and }  V_5= 1.055 \text{ p.u.}$, as they have shown the best cumulative reward value among several other configurations.

\subsection{Results}
In this section, we assess the performance of the proposed RRL model through a comprehensive evaluation. Initially, we compare the learning progress and overall cumulative rewards of the RRL framework against the baseline SDC controller and pure RL model. Furthermore, we evaluate the models based on key performance metrics, including the number of voltage violations resolved in the test dataset, the steps required to resolve voltage violations, and the percentage of power utilization, which indicates the extent of active power curtailment applied. Finally, an ablation study is conducted to highlight the significance of the proposed architecture.
\subsubsection{Learning Process}
We show the learning curves of different approaches by plotting the evolution of cumulative episode reward in Figure \ref{fig:learning-curves}.
The comparison includes the RRL approach with different weighting coefficients $\lambda\in\{1.5, 1.0, 0.5\}$, the vanilla RL approach, and the SDC control prior.
The results indicate that the pure RL model exhibits the slowest convergence, reaching the cumulative reward level of the baseline SDC controller only after approximately 700,000 training steps. In contrast, all RRL-based models surpass the SDC controller's performance within the first 100,000 training steps, highlighting the advantages of residual policy learning in accelerating convergence and enhancing exploration efficiency.
Further, we can see how the choice of $\lambda$ influences the training procedure: with increasing weight, the RL agent gains more agency over the combined action, can explore a broader action space, and thus learns a better policy. However, the improvements due to the weight increase saturate, with 1.0-RRL and 1.5-RRL having comparable asymptotic performance. The downside of higher exploration is a higher variance and a slow start in training, which manifests as the initial dip in evaluation cumulative reward.
This tradeoff can be reduced by employing adaptive weighting approaches\cite{cramer2024contextualized}.
\input{Figures/Learning-Cruves}
\subsubsection{Overall Performance}
Table \ref{table:result-metrics} provides a comparison of the overall performance of the proposed 1.5-RRL model, the pure RL approach, and the baseline SDC controller on the test dataset.
The results are consistent with the observed learning curves and further validate the effectiveness of the proposed RRL framework. The latter not only effectively mitigates voltage violations but also achieves the highest power usage percentage, demonstrating its ability to prioritize reactive power provision while minimizing active power curtailment. These results emphasize the efficiency of the proposed framework in balancing voltage regulation and power curtailment.
\input{Figures/Result-Table}

Figure \ref{fig:voltage-and-power-episode} illustrates the evolution of the maximum voltage throughout a representative episode. At step 0, we can see that an overvoltage is present, as the voltage surpasses the $1.05$ p.u. limit. All methods successfully bring the voltage within safe limits within the initial steps. However, the RRL model demonstrates superior performance by achieving voltage regulation with fewer control steps and minimal active power curtailment. This highlights its effectiveness in ensuring near-optimal power utilization while efficiently resolving voltage violations.
\input{Figures/Voltage-power-over-episode}
\subsubsection{Ablation of Network Architecture and SDC controller Observation}
To investigate the importance of our design choices regarding the Q-network architecture (Section \ref{sec:network-architecture}) and observation of SDC controller actions (Section \ref{sec:observing-controller-actions}), we compare four configurations: Q-network with (CNNQ) or without (MLPQ) prepended shared local linear layers, each with (CTRL) or without controller action observation. Learning curves for all variants are plotted in Figure~\ref{fig:ablation-study}. All variants use the RRL approach with $\lambda=1.5$ and the same actor network architecture. Variant CNNQ + CTRL corresponds to the 1.5-RRL presented earlier.
\input{Figures/Ablation-Learning-Curves}

The results show that using our proposed network architecture (CNNQ) significantly improves training stability and enhances performance compared to its counterpart (MLPQ) in both cases. Notably, while the inclusion of controller actions in the observation space leads to an increase in asymptotic performance for the CNN Q-network, it causes failure without it. One possible explanation is the large increase in observation dimensionality, which cannot be handled gracefully by the simple MLP architecture.

These findings emphasize the effectiveness of the proposed CNN-like Q-network architecture in combination with the observation of SDC controller actions for improving learning efficiency, convergence stability, and asymptotic performance.

%% file: Figures/Grid.tex
\begin{figure}[ht!]
    \centering
    \includegraphics[scale=0.8]{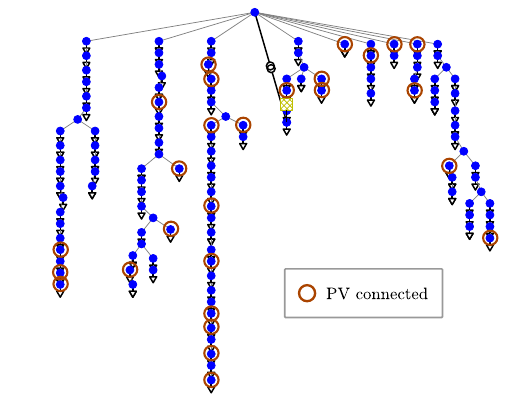}
    \caption{Distribution grid used in training and evaluation.}
    \label{fig:electrical-grid}
\end{figure}

%% file: Figures/Hyperparameter.tex
\begin{table}[htbp]
\centering
\caption{Hyperparameters}

\begin{tabular}{|l|l|c|}
\hline
 & \textbf{Hyperparameter} & \textbf{Value} \\ \hline
\textbf{Q-net (Critic)} & Hidden local layers & \num{2} \\ 
                       & Hidden local layer size & \num{32} \\ 
                       & Hidden global layers & \num{2} \\ 
                       & Hidden global layer size & \num{512} \\ \hline
\textbf{Actor}         & Attention blocks & \num{4} \\ 
                       & Model dimension & \num{32} \\ 
                       & Attention heads & \num{4} \\ 
                       & Feedforward layer size & \num{64} \\ \hline
\textbf{Training}      & Batch size & \num{256} \\
                       & Learning rate (Actor \& Q-net) & \num{3e-4} \\
                       & Replay buffer size & \num{1e6} \\
                       & Discount factor $\gamma$ & \num{0.99} \\ 
                       & SAC target entropy & $-\text{dim}(\A) = -54$ \\ \hline
                       
\end{tabular}
\label{tab:hyperparameters}
\end{table}

%% file: Figures/Learning-Cruves.tex
\begin{figure}[ht!]
    \centering
    \includegraphics{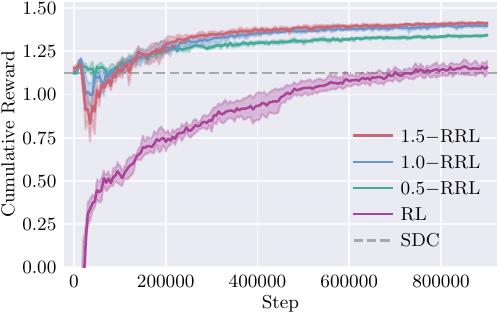}
    \caption{Evaluation Learning curves of RRL and pure RL approach.}
    \label{fig:learning-curves}
\end{figure}

%% file: Figures/Result-Table.tex
\begin{table}[htbp]
    \centering
    \caption{Performance Metrics for Different RL Algorithms}
    \resizebox{\columnwidth}{!}{
    \begin{tabular}{lccc}
         & Violations Solved [\%] & Steps to Solve & Power Usage [\%] \\
        \hline
        \\[-1.5ex]  
        $1.5$-RRL & \num{99.9} $[\num{99.8}, \num{100}]$  & \num{1.62} $[\num{1.60}, \num{1.63}]$ & \num{95.2} $[\num{94.8}, \num{95.6}]$ \\[1ex] 
        RL & \num{99.6} $[\num{99.5}, \num{99.8}]$ & \num{1.68} $[\num{1.67}, \num{1.70}]$ & \num{72.7} $[\num{70.5}, \num{74.5}]$ \\[1ex] 
        SDC & \num{95.6} & \num{1.98} & \num{79.3} \\[0.5ex] 
        \hline
    \end{tabular}}
    \label{table:result-metrics}
\end{table}

%% file: Figures/Voltage-power-over-episode.tex
\begin{figure}[ht!]
    \centering
    \includegraphics{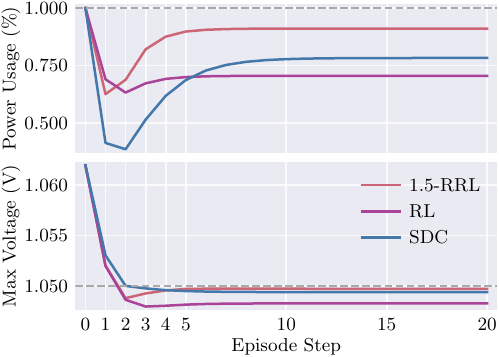}
    \caption{Voltage and power usage over one episode}
    \label{fig:voltage-and-power-episode}
\end{figure}

%% file: Figures/Ablation-Learning-Curves.tex
\begin{figure}[ht!]
    \centering
    \includegraphics{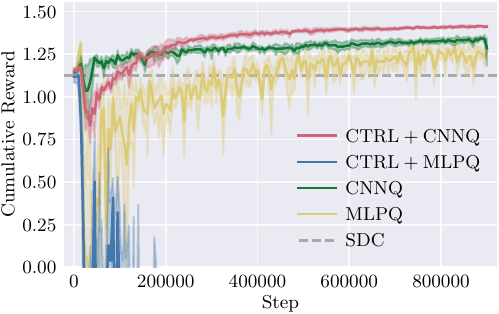}
    \caption{Evaluation learning curves for ablation study.}
    \label{fig:ablation-study}
\end{figure}

%% file: 06_Conclusion.tex
\subsection{Conclusion}
This paper introduces a novel Residual Reinforcement Learning (RRL) framework for voltage control in distribution grids, which enhances control performance by learning a residual action on top of an existing control prior—the modified Sequential Droop Control (SDC) mechanism. A key contribution of our approach is its ability to efficiently control both active and reactive power of PV-inverters without requiring full state information of the power grid, thereby offering a more practical and scalable solution.

Our experimental evaluation demonstrates that the proposed RRL framework outperforms both the baseline SDC controller and pure RL models, effectively resolving voltage violations while significantly minimizing active power curtailment. Another major contribution of this work is the introduction of a new neural network architecture, featuring local shared linear layers for the Q-network and a Transformer-Encoder for the actor network. This architecture has shown to significantly enhance convergence speed and overall performance, especially when incorporating the controller's action in the observation space of the RL agent.